\documentclass[traditabstract]{aa}
\usepackage{graphicx}
\usepackage{txfonts}

 \newcommand{\mic}{$\mu$m}
 \newcommand{\mics}{$\mu$m~}

\def\tex {\ifmmode{{T}_{\rm ex}}\else{$T_{\rm ex}$}\fi}
\def\tmb {\ifmmode{{T}_{\rm mb}}\else{$T_{\rm mb}$}\fi}
\def\ci     {\ifmmode{{\rm C}{\rm \small I}}\else{C\ts {\scriptsize I}}\fi}
\def\hi     {\ifmmode{{\rm H}{\rm \small I}}\else{H\ts {\scriptsize I}}\fi}
\def\hh     {\ifmmode{{\rm H}_2}\else{H$_2$}\fi}

\def\ts     {\thinspace}
\def\kms    {\ifmmode{{\rm \ts km\ts s}^{-1}}\else{\ts km\ts s$^{-1}$}\fi}
\def\msol   {\ifmmode{{\rm M}_{\odot}}\else{M$_{\odot}$}\fi}
\def\lsol   {\ifmmode{{\rm L}_{\odot}}\else{L$_{\odot}$}\fi}
\def\zsol   {\ifmmode{{\rm Z}_{\odot}}\else{Z$_{\odot}$}\fi}
\def\etal   {{\rm et\ts al.\ts}}

\begin{document}

\title{Dust and gas power-spectrum in M33 (HERM33ES)\thanks{{\em Herschel} is an ESA space observatory with science instruments provided by European--led Principal Investigator consortia and with important participation from NASA.}}
   \author{F. Combes\inst{1} \and M. Boquien\inst{2} \and C. Kramer\inst{3} \and E. M. Xilouris\inst{4} \and F. Bertoldi\inst{5} \and J. Braine\inst{6} \and C. Buchbender\inst{3} \and D. Calzetti\inst{7} \and P. Gratier\inst{8} \and F. Israel\inst{9} \and B. Koribalski\inst{10} \and S. Lord\inst{11} \and G. Quintana-Lacaci\inst{3} \and M. Rela\~{n}o\inst{12} \and M. R\"ollig\inst{13} \and G. Stacey\inst{14} \and F. S. Tabatabaei\inst{15} \and R. P. J. Tilanus\inst{16} \and F. van der Tak\inst{17} \and P. van der Werf\inst{9} \and S. Verley\inst{12}}

\offprints{F. Combes}
\institute{Observatoire de Paris, LERMA (CNRS:UMR8112), 61 Av. de l'Observatoire, F-75014, Paris, France
\email{francoise.combes@obspm.fr}
   \and Laboratoire d'Astrophysique de Marseille, UMR 6110 CNRS, 38 rue F. Joliot-Curie, 13388, Marseille, France
   \and Instituto Radioastronomia Milimetrica, Av. Divina Pastora 7, Nucleo Central, E-18012 Granada, Spain
   \and Institute of Astronomy and Astrophysics, National Observatory of Athens, P. Penteli, 15236 Athens, Greece
   \and Argelander Institut f\"ur Astronomie. Auf dem H\"ugel 71, D-53121 Bonn, Germany
   \and Laboratoire d'Astrophysique de Bordeaux, Universit\'{e} Bordeaux 1, Observatoire de Bordeaux, OASU, UMR 5804, CNRS/INSU, B.P. 89, Floirac F-33270, France
   \and University of Massachusetts, Department of Astronomy, LGRT-B 619E, Amherst, MA 01003, USA
   \and IRAM-Institut de Radio Astronomie Millim\'etrique, 300 Rue de la Piscine, 38406-St. Martin d`H\`eres, France
   \and Leiden Observatory, Leiden University, PO Box 9513, NL 2300 RA Leiden, The Netherlands
   \and ATNF, CSIRO, PO Box 76, Epping, NSW 1710, Australia
   \and IPAC, MS 100-22 California Institute of Technology, Pasadena, CA 91125, USA
   \and Dept. F\'{i}sica Te\'{o}rica y del Cosmos, Universidad de Granada, Spain
   \and KOSMA, I. Physikalisches Institut, Universit\"at zu K\"oln, Z\"ulpicher Stra\ss{}e 77, D-50937 K\"oln, Germany
   \and Department of Astronomy, Cornell University, Ithaca, NY 14853, USA
   \and Max Planck Institut f\"ur Astronomie, K\"onigstuhl 17, D-69117 Heidelberg, Germany
   \and JAC, 660 North A'ohoku Place, University Park, Hilo, HI 96720, USA
   \and SRON Netherlands Institute for Space Research, Landleven 12, 9747 AD Groningen, The Netherlands
}

   \date{Received 16 October 2011/ Accepted 10 January 2012}

   \titlerunning{Dust power-spectrum in M33}
   \authorrunning{F. Combes et al.}

   \abstract{Power spectra of de-projected images of late-type galaxies in gas and/or dust emission are very useful
diagnostics of the dynamics and stability of their interstellar medium. Previous studies have shown
that the power spectra can be approximated as two power-laws, a shallow one at large scale 
(larger than 500 pc) and a steeper
one at small scale, with the break between the two corresponding to the line-of-sight thickness of the
galaxy disk. The break separates the 3D behaviour of the interstellar medium
at small scale, controlled by star formation and feedback, from the 2D behaviour at large scale,
driven by density waves in the disk. 
The break between these two regimes depends on the thickness
 of the plane which is determined by the natural     self-gravitating scale of the interstellar medium.
We present a thorough analysis of the power
spectra of the dust and gas emission at several wavelengths in the nearby galaxy M33.
In particular, we use the recently obtained images at five wavelengths by PACS and SPIRE onboard {\em Herschel}.
The large dynamical range (2-3 dex in scale) of most images allows us to determine
clearly the change in slopes from -1.5 to -4, with some variations with wavelength.
The break scale is increasing with wavelength, from 100 pc at 24 and 100\mics to 350 pc at 500\mic, 
suggesting that the cool dust lies in a thicker disk than the warm dust, may be due to star formation more 
confined to the plane. The slope at small scale tends to be steeper at longer wavelength,
meaning that the warmer dust is more concentrated in clumps.
Numerical simulations of an isolated late-type galaxy, rich in gas and with no bulge, like M33,
are carried out, in order to better interpret these observed results. Varying the star formation and feedback
parameters, it is possible to obtain a range of power-spectra, with two power-law slopes
and breaks, which nicely bracket the data. The small-scale power-law is indeed reflecting the 3D behaviour
of the gas layer, steepening strongly while the feedback smoothes the structures, by increasing the gas turbulence. M33 appears to correspond to a fiducial model with an SFR of $\sim$ 0.7 M$_\odot$/yr, 
with 10\% supernovae energy coupled to the gas kinematics. 
}
\maketitle

%---------------------------------------------------------------

\section{Introduction}

  Quantifying the multi-scale structure of the interstellar medium (ISM) in galaxies is a difficult
enterprise, but rewarding since it betrays the underlying physical phenomena,
controlling its dynamics. It has been known for a long time that the ISM reveals no
preferential scale and is well represented by a fractal structure, or a power-law power-spectrum.
 To study the atomic ISM, Crovisier \& Dickey (1983) and Green (1993) studied the power-spectrum as a function 
of inverse scale of the 21cm emission, and found a power-law of slope between -2 and -3, 
independent of the distance and the velocity of the HI. This power-law slope was observed to be
steeper from -3 to -4, when the broadest velocities were averaged, thus including the warm
gas, with more turbulence (Dickey \etal 2001). Goldman (2000) and Stanimirovic \etal (2000)
extended the ISM studies in the Milky Way to the Small Magellanic Cloud (SMC), where
a power-law of slope -3.1 to -3.4 was found for the dust and gas, and interpreted as 3D turbulence,
although the index is shallower than the -3.7 Kolmogorov spectrum (Stanimirovic \& Lazarian 2001).
  A break in scale has been searched for in order to identify the energy injection mechanism,
but none has been found at small scale (i.e. smaller than 500 pc). Instead a flattening 
towards larger scales has been measured by
Elmegreen \etal (2001) in the atomic gas of the Large Magellanic Cloud (LMC). They interpret this flattening
as a change of dimensionality, from 3D at small scale to 2D at large scale, where the dynamics
of the galaxy plane is dominating, and identify the scale of the break as the thickness of the plane.
This method to find the scale height of disks has been developed by Padoan \etal (2001), who
found $\sim $ 180 pc in the LMC. Recently, Block \etal (2010) have analysed the dust 
maps of the LMC from Spitzer, and also confirm this break scale at 100-200 pc. From high resolution
numerical simulations fitted to a late-type galaxy like the LMC, Bournaud \etal (2010) were
able to retrieve the characteristics of the observed dust power-spectrum.
The source of turbulence could then be due to large-scale differential rotation,
cold gas accretion or galaxy interactions (Lazarian \etal 2001, Elmegreen \& Scalo 2004).
 Khalil \etal (2006), from a 2D wavelet analysis of HI in our own Galaxy,
find a power-law slope of -3 everywhere over the second quadrant, and
 do not detect the variations with velocity slices predicted by Lazarian \& Pogosyan (2004). 
However they discover an anisotropy linked to spiral arms. The fractal approach was widely
developed for molecular clouds, which are more self-gravitating than the HI gas (Falgarone \etal 1991,
Pfenniger \& Combes 1994, Stutzki \etal 1998).

 Up to now, the break scale in the HI distribution has not been identified unambiguously
in external galaxies other than the LMC, although it has been searched for.
Dutta \etal (2008) studied the face-on galaxy NGC 628, and found an HI power spectrum
with a single power-law slope of -1.6, over the scales 0.8 to 8 kpc, their spatial resolution
being insufficient to reach the scale height of the gaseous plane.
In DDO210, Begum \etal (2006) found a power-law slope of -2.75  over the scales 80  to 500 pc.
This steep slope is compatible with what is found in the MW and LMC/SMC at small scales,
driven by 3D-turbulence. In other dwarfs also (Dutta \etal 2009b) a single slope
is found. The face-on galaxy NGC1058 is the first one for which the 
transition from 3D to 2D turbulence has been claimed (Dutta \etal 2009a): the slope
changes from -2.5 over 0.6 to 1.5 kpc, to -1 over 1.5-10 kpc scales. The scale height is
then estimated at 490 pc, which is a relatively large value for the HI layer.
 In the SMC, while a steep power-law is found consistently by several studies,
no break scale has been found, and no transition to the 2D-turbulence (Roychowdhury \etal 2010).
It is suggested that dwarf galaxies have gas disks considerably thicker than normal spiral
galaxies, which support previous studies from the thickness of dust lanes
in edge-on galaxies (e.g. Dalcanton \etal 2004). The transition between thick and thin gas disks
was found at a mass corresponding to a circular velocity of 120 \kms.
 For galaxies with low inclination on the sky plane, the thickness of 
gas disks can be estimated through the ratio of gas velocity dispersion to rotation
(Dalcanton \& Stilp 2010).
 The case of M33 is exactly intermediate, having a rotational velocity of 120 \kms. Both thick and
thin disks could be expected. 

Another tool to explore the scale height of disks with higher spatial resolution, 
apart from studying edge-on galaxies (e.g. Bianchi \& Xilouris 2011),
is to study young components related to the gas, as young stars just formed out of the ISM.
Elmegreen \etal (2003a) studied the power-spectrum of star formation maps in nearby galaxies,
and found the same behaviour as for the LMC in HI: a steep power-law at small scales, 
typical of 3D-turbulence, with a flattening at large scales, and a break corresponding to
several hundred parsecs, typical of the plane thickness. This is true for flocculent and
density-wave galaxies, although the spiral arm wave dominates at very large scales.
In M33, Elmegreen \etal (2003b) analysed optical and H$\alpha$ images and found
a very shallow power-law of slope -0.7, which they attribute to flattening due to contamination
by strong point sources (foreground stars). For the face-on galaxy NGC 628, a shallow slope
of -1.5 is found, with no break in spite of the very high spatial resolution obtained with HST 
(scale below 10 pc, Elmegreen \etal 2006). 
Odekon (2008) however finds a transition scale of about 300 pc between a steep spectrum
at small scale and flattening at large scale, in the stellar distribution of M33. This scale
does not change much with the age of the stellar population considered, however the slope is steeper
for the bluest and youngest stars.
Sanchez \etal (2010) investigate all the young components in M33, including young stars, HII regions
and molecular clouds, and also find a clear transition region, but in the range 0.5-1 kpc. The fact that
 different results are obtained for the same galaxies calls for new studies to clarify the situation.

The galaxy M33 has been observed with {\em Herschel} PACS and SPIRE at all wavelengths, 
by the Key Project HERM33ES,
and the photometric maps have been presented in Kramer \etal (2010), Boquien \etal (2010) and
Verley \etal (2010). These provide high signal-to-noise maps with a large dynamical range of scales
of the dust emission at many wavelength, allowing to probe warm and cool dust. 
The large and small scale structures of the gas and dust are studied through Fourier analysis.
 We present in Section \ref{obs} the different maps used, and how they are processed,
and deprojected to a face-on orientation. The results will be compared
to numerical models, which technique is described in Section \ref{model}. Results are presented
for both observations and models in Section \ref{res}, and discussed in Section \ref{disc}.
 We adopt a distance of 840 kpc for M33  (Freedman \etal 1991).

\begin{figure}[htp]
\centering
\includegraphics[angle=-90,width=6cm]{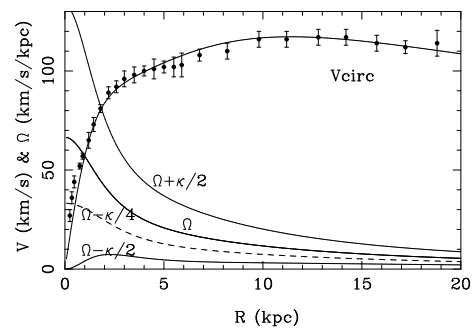}
\caption{Rotation curve for our late-type Scd model, compared to 
the observed points (symbols and error bars) compiled for M33 by
Corbelli \& Salucci (2007). In addition to the circular velocity
of the model, the characteristic frequencies
$\Omega$, $\Omega-\kappa/2$, etc. are plotted.}
\label{vcur}
\end{figure}

%--------------------------------------Five column Table-----------------1-
\begin{table}[h]
      \caption[]{Characteristics of the different maps used}
         \label{tab:maps}
            \begin{tabular}{l c c c c c}
            \hline
            \noalign{\smallskip}
  Map & \multicolumn {2}{c}{Resolution}   & Npix & \multicolumn{2}{c}{pixel size} \\
          & (\arcsec) & (pc)  &      & (\arcsec) & (pc) \\
            \noalign{\smallskip}
            \hline
            \noalign{\smallskip}
Galex-FUV & 4.3'' & 17 &  2560 & 1.5'' & 6\\
Galex-NUV & 5.3'' & 21 &  2560 & 1.5'' & 6\\
H$\alpha$ & 6'' & 24&  1920 & 2'' & 8\\
MIPS-24\mics & 6'' & 24 &  2560 & 1.5'' & 6\\
MIPS-70\mics & 18'' & 72 &  850 & 4.5'' & 18\\
PACS-100\mics & 7'' & 28&  2268 & 1.7'' & 7\\
PACS-160\mics & 11'' & 44 &  1344 & 2.85'' & 12\\
SPIRE-250\mics & 18'' & 72&   800 & 6'' & 24\\
SPIRE-350\mics & 26'' & 104 &   480 & 10'' & 41\\
SPIRE-500\mics & 36'' & 144 &   342 & 14'' & 57\\
CO(2-1) & 12''& 48   & 1470  & 2'' & 8\\
HI-21cm & 12''& 48  & 1200  & 4'' & 16\\
%HI-21cm & 5'' & 20 & 3240  & 1.5'' & 6\\
            \noalign{\smallskip}
            \hline
           \end{tabular}
\\A distance of 840 kpc has been adopted for M33. 1'' = 4 pc
\\Resolutions correspond to the major axis direction, 
\\they are 1/cos(56$^\circ$) = 1.79 larger in the other direction
\\ Npix is the number of pixels in each dimension of the map
\end{table}

\begin{figure}[htp]
\includegraphics[angle=-90,width=7cm,clip=true]{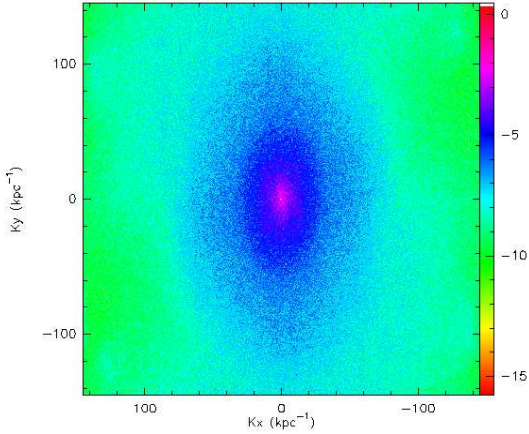}
\caption{2D power-spectrum for the PACS 100\mics image. The color scale
is in arbitrary units.}
\label{fig:2D}
\end{figure}

\begin{figure}[htp]
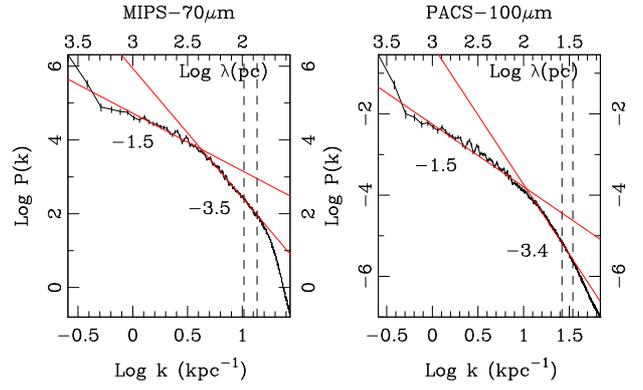

\includegraphics[angle=-90,width=4cm]{18282f3a.ps}
\includegraphics[angle=-90,width=4cm]{18282f3b.ps}
\caption{Power spectrum of the Spitzer-MIPS 70\mics map,
and the PACS-100\mics dust emission map of M33.  The vertical dash line at right represents
the spatial resolution of the observations, on the major axis.
 Since the de-projection to a face-on galaxy implies a factor 1/cos(56$^\circ$) = 1.79 larger beam on
the minor axis, we have also plotted a second dashed vertical line at left,
indicating the geometric mean of the resolution on both minor and major axis.
Statistical error bars are given in this first figure, and then omitted later for clarity.
They are significant only at the largest scales (small $k$), where the statistical errors are
the largest.}
\label{fig-pow1}
\end{figure}

\section{Observational data}
\label{obs}

To study the interstellar medium structure, we use
all possible tracers, beginning by gas and dust emission
(HI, CO lines, and far-infrared photometry), and also
star formation from the gas (H$\alpha$, ultra-violet continuum).

 For the dust emission, we use {\em Herschel} observations taken in the
context of the HERM33ES open time key project
(Kramer \etal 2010) in combination with Spitzer
MIPS data (Verley \etal 2007),
spanning wavelengths from 24\mics to 500\mic.

The {\em Herschel} PACS data covering the
100\mics  and 160\mics  bands, have been described in
Boquien \etal (2010, 2011). The SPIRE observations obtained
at 250, 350 and 500\mics are displayed in 
Kramer \etal (2010) and Verley \etal (2010).

We also use ground-based H$\alpha$ observations presented
in Hoopes \etal (2001) in order to
trace the current ($\sim$ 10 Myr) star formation and
GALEX data (Thilker \etal 2005, Gil de Paz \etal 2007) to trace the
recent ($\sim$ 100 Myr) star formation. To trace the atomic
gas, we use the HI 12\arcsec\ts resolution map published in Gratier \etal (2010),
and for the molecular gas, their partial CO(2-1)  12\arcsec\ts resolution map
(Gratier \etal 2010).

All useful characteristics of the images are summarized in Table \ref{tab:maps}.
All maps have been deprojected, with the adopted inclination of 56$^\circ$ and position
angle of 22.5$^\circ$, before Fourier analysis.

The data analysed in the present work are more
appropriate than earlier data (Elmegreen \etal 2003b, Odekon 2008, Sanchez \etal 2010) 
to study the power spectra 
and the structure of the interstellar medium, since they refer directly to the gas and dust
in M33, and they have the dynamical range (from 40 pc to 20 kpc) and spatial resolution to 
reveal the thickness of the gas layer.  The gas and dust emission trace directly
the collisional material, with no intervening and perturbing emission (like galactic stars).

\begin{figure}[htp]
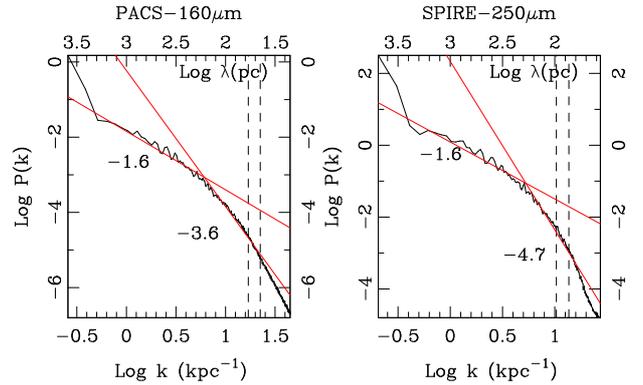

\includegraphics[angle=-90,width=4cm]{18282f4a.ps}
\includegraphics[angle=-90,width=4cm]{18282f4b.ps}
\caption{Same as Figure \ref{fig-pow1}, for the power spectrum of the PACS-160\mics and SPIRE-250\mics dust map. }
\label{fig-pow2}
\end{figure}

\begin{figure}[htp]
\includegraphics[angle=-90,width=4cm]{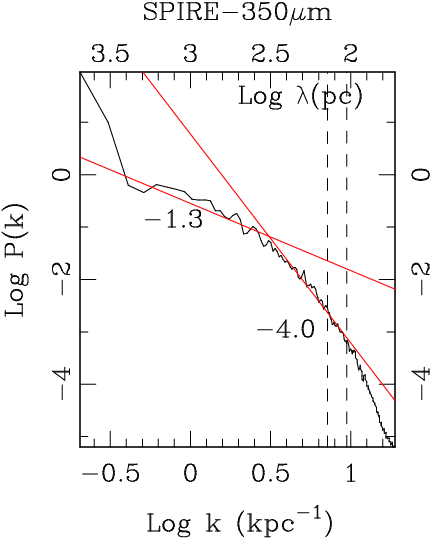}
\includegraphics[angle=-90,width=4cm]{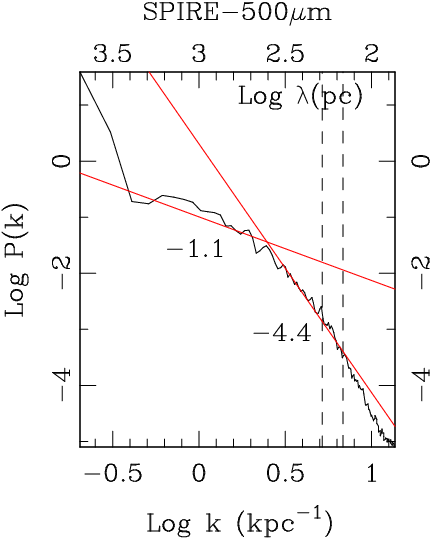}
\caption{Same as Figure \ref{fig-pow1}, for the power spectrum of the SPIRE-350\mics and 500\mics cool dust map. }
\label{fig-pow3}
\end{figure}

\section{Numerical model}
\label{model}

In order to compare with what is expected from the dynamical instabilities of the 
interstellar medium, we have run idealised simulations, fitted for a late-type galaxy without 
bulge and rich in gas like M33.

\subsection{Technique}
The numerical code is based on the TREE/SPH version developed by Semelin \& Combes (2005)
and di Matteo \etal (2007), with star formation and feedback.
The TREE gravitational part of the code deals with two collisionless components: the stars and
the dark matter, which differ only by their initial conditions. It deals also with the gas
component, whose hydrodynamics is treated with the SPH (Smooth Particle Hydrodynamics)
 algorithm, using adaptive resolution.
 The softening parameter, or the spatial resolution of gravity, 
is taken to be 30 pc, and the average size of a gaseous
particle (the SPH kernel, or spatial resolution of the hydrodynamics) 
is also around 30 pc, although it can drop down to 3 pc in overdense regions.
  The number of particles is 1.2 million, with 0.6 million gas particles, 0.4 million
dark matter particles, and 0.2 million "stars" at the beginning of the simulations.
 The gas particles are assumed to follow an isothermal equation of state, with pressure
forces corresponding to a sound velocity of 6 \kms, in the fiducial model.
Shocks are treated with a conventional form of artificial viscosity,
with parameters given in di Matteo \etal (2007).
All physical quantities (density and forces) are computed averaging
over a number of 50 neighbours. 
The equations of motion are integrated using a leapfrog algorithm
with a fixed time step of 0.3 Myr. 

 Star Formation is taken into account in most models, with an assumed Schmidt law,
of exponent n=1.5, i.e. for each gas particle of mass $M_{gas}$, and volumic density
$\rho_{gas}$ computed over its neighbours:
$$
\frac{dM_{gas}}{M_{gas}dt} = C_* \,  \rho_{gas}^{1/2}
$$
\noindent where in most runs, the star formation constant $C_*$ is calibrated such 
that the gas consumption time-scale is of the order of  2Gyr (then $C_* = C_0$). 
 $C_*$ is increased in some of the runs, to check its influence on the gas scale height.
 In most simulations, the density threshold for star formation is set very low,
and is then increased in a few runs.
 Star formation is then taken into account for each gas particle, using the 
hybrid particles scheme (e.g. di Matteo \etal 2007), before sufficient stellar mass is accumulated
and a star particle can be created. Continuous mass loss is also considered, with 
about 40\% of the stellar mass loss by young massive stars after 5 Gyr (Jungwiert \etal 2001).

  The energy reinjected  by supernovae into the interstellar medium is assumed
to be a kinematic feedback on neighboring gas particles: each neighbour of a SPH particle
having formed $\delta$m of stars is given a radial kick in velocity away from
the supernova formed. The energy available is computed from $\delta$m, assuming 
a Scalo IMF (about 0.5\% of the stars formed have a mass larger than 8 M$_\odot$ and 
explode as supernovae), the energy of a supernova $E_{SN} = 10^{51}$ erg is distributed
to the 50 neighbours according to their weight (the SPH kernel), with an efficiency $\epsilon$, which has
been varied from $10^{-4}$ to 1, with 10\% being the fiducial value (see Table \ref{tab:CI}).
For the fiducial feedback efficiency, a particle might get a kick of $\sim$ 1 \kms at each time-step,
for the extreme feedback, it can reach 5 \kms. The efficiency of the feedback is not well known,
and depends on the coupling of the supernovae and the gas (whether the young stars have
already left their birth clouds when they explode, etc..). Some energy and supernova explosions
 are also due to the old population of stars (e.g. Dopita 1985).

\subsection{Initial conditions}

 The initial stellar component of the galaxy is selected to be of type Scd,
similar to the M33 galaxy.

The stellar and gaseous disks are represented
initially by Miyamoto-Nagai functions, with characteristic
height parameters of 170 pc and 70 pc respectively.
The stellar disk has a mass of 0.5 $\times$ 10$^{10}$ M$_\odot$, with 
a characteristic radius of 2 kpc, while the gas disk has a 
mass of 1.5 $\times$ 10$^{9}$ M$_\odot$, with a radius 2.5 kpc.
Both are embedded in a spherical halo of dark matter,
represented by a Plummer sphere of mass 
4  $\times$ 10$^{10}$ M$_\odot$ within 18 kpc, with a characteristic radius of
9.5 kpc.

The initial stability of the gas and stellar disks is controlled 
by the Toomre parameter Q, ratio of the radial component
of the velocity dispersion to the critical one 
$\sigma_{crit}= 3.36 G \Sigma /\kappa$, $\Sigma$
being the disk surface density of the component (gas or stars),
and $\kappa$ the epicyclic frequency.  We select equal Toomre
parameters for gaseous and stellar disks.
The azimuthal velocity dispersions verify initially the
epicyclic ratio $\sigma_\theta /\sigma_r$ = $\kappa / 2\Omega$,
and the z-components are fixed by the isothermal equilibrium
of the disks, in the vertical direction.

The gas is isothermal, but its equivalent velocity dispersion,
governing the pressure, may be varied 
between 6 and 12 \kms. The recipes for star formation: efficiency,
density threshold, and feedback, are also varied; they
have a significant impact on the interstellar gas structure.

The initial conditions  of the runs
described here are given in Table \ref{tab:CI}.
The rotation curve corresponding to one of the runs
is plotted in comparison to the data points in Figure \ref{vcur}.
All runs have similar rotation curves.
Given the uncertainties, all the rotation curves are compatible with the data.

%--------------------------------------Six column Table-----------------1-
\begin{table}[h]
      \caption[]{Initial conditions of the simulations}
         \label{tab:CI}
            \begin{tabular}{l c c c cc}
            \hline
            \noalign{\smallskip}
  Run & $Q_*$   & $V_s$ & $C_*$ & threshold &$\epsilon$ \\
     & =$Q_g$   &   (km/s) &    &        at. cm$^{-3}$      &   (feedback) \\
            \noalign{\smallskip}
            \hline
            \noalign{\smallskip}
run1 & 2. &  6. & $C_0$  & 2 $\times$ 10$^{-6}$  &$10^{-4}$\\
run2 & 2. &  6. & $C_0$  &  2 $\times$ 10$^{-6}$ &$10^{-3}$\\
run3 & 2. &  6. & $C_0$  & 2 $\times$ 10$^{-6}$  & 1. \\
run4 & 2. &  6. & $C_0$  & 2 $\times$ 10$^{-6}$  & $10^{-1}$\\
run5 & 1.5 &  6. & $C_0$x20  & 2 $\times$ 10$^{-3}$  & $10^{-1}$\\
run6 & 1. &  6. & $C_0$x20  & 4   & $10^{-1}$\\
run7 & 1. &  6. &   0.  &  -- &  --\\
run8 & 1. &  12. & 0.  &  -- & --\\
run9 & 1. &  12. & $C_0$  &  2 $\times$ 10$^{-6}$ &  $10^{-1}$\\
            \noalign{\smallskip}
            \hline
           \end{tabular}
\\ $C_0$ is such that the gas depletion time is 2Gyr
\end{table}

\section{Results}
\label{res}

\subsection{Power-spectrum of the data}
\label{pow-m33}

The power spectra of the 2D projected maps were analysed through the 
2D Fourier transform:

$$
{F^*}(k_x,k_y)=\int_x \int_y F(x,y)e^{-i(k_xx + k_yy)}dxdy
$$
%\end{equation}

\noindent where $F(x,y)$ is the intensity of each pixel, and $k_x$ and $k_y$ are the
wave number, conjugate  to $x$ and $y$, varying as the inverse of scales. The full 2D power
spectrum is given by

$$
P(k_x,k_y) = (Re {[F^*]})^2 + (Im {[F^*]})^2
$$

\begin{figure}[htp]
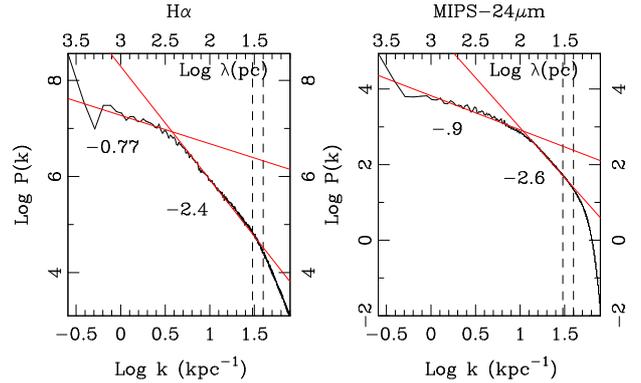

\includegraphics[angle=-90,width=4cm]{18282f6a.ps}
\includegraphics[angle=-90,width=4cm]{18282f6b.ps}
\caption{Same as Figure \ref{fig-pow1}, for the power spectrum of the H$\alpha$ and MIPS-24\mics maps. }
\label{fig-pow4}
\end{figure}

\begin{figure}[htp]
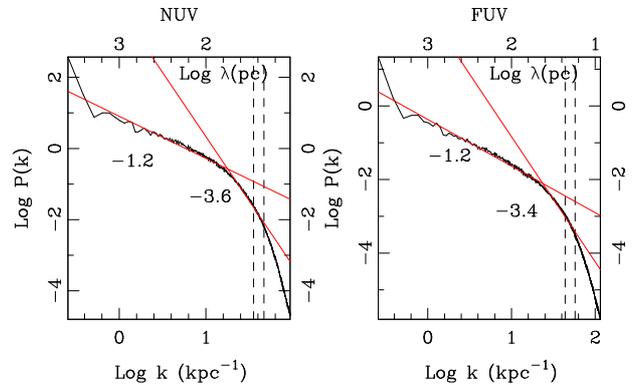

\includegraphics[angle=-90,width=4cm]{18282f7a.ps}
\includegraphics[angle=-90,width=4cm]{18282f7b.ps}
\caption{Same as Figure \ref{fig-pow1}, for the power spectrum of the GALEX NUV and FUV maps. }
\label{fig-pow5}
\end{figure}

\begin{figure}[htp]
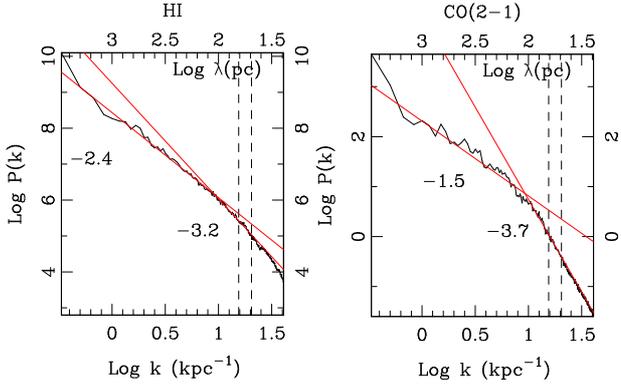

\includegraphics[angle=-90,width=4cm]{18282f8a.ps}
\includegraphics[angle=-90,width=4cm]{18282f8b.ps}
\caption{Same as Figure \ref{fig-pow1}, for the power spectrum of the HI and CO(2-1) maps. }
\label{fig-pow6}
\end{figure}

\begin{figure*}
\includegraphics[angle=-90,width=8.4cm,clip=true]{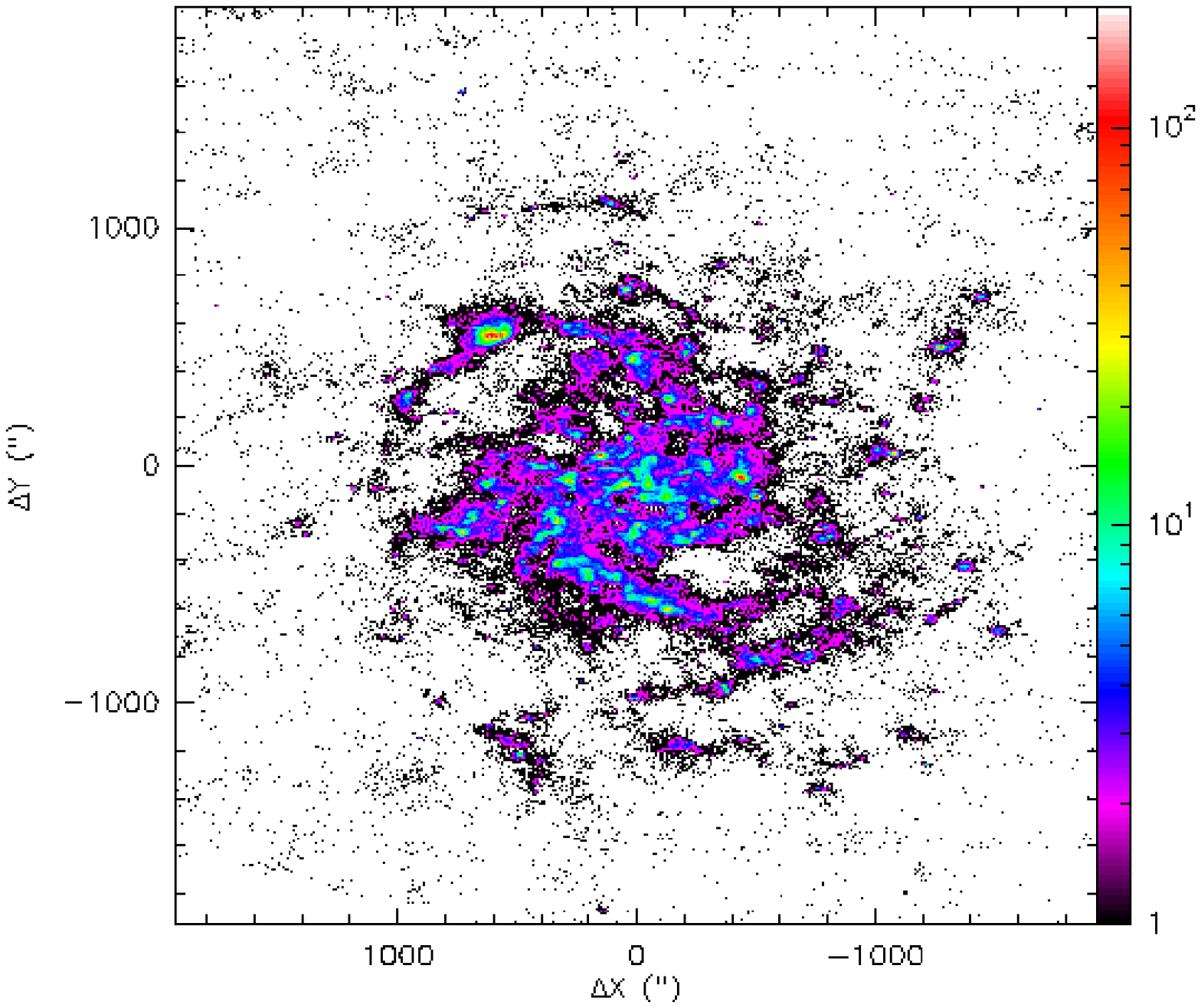}
\includegraphics[angle=-90,width=8.4cm,clip=true]{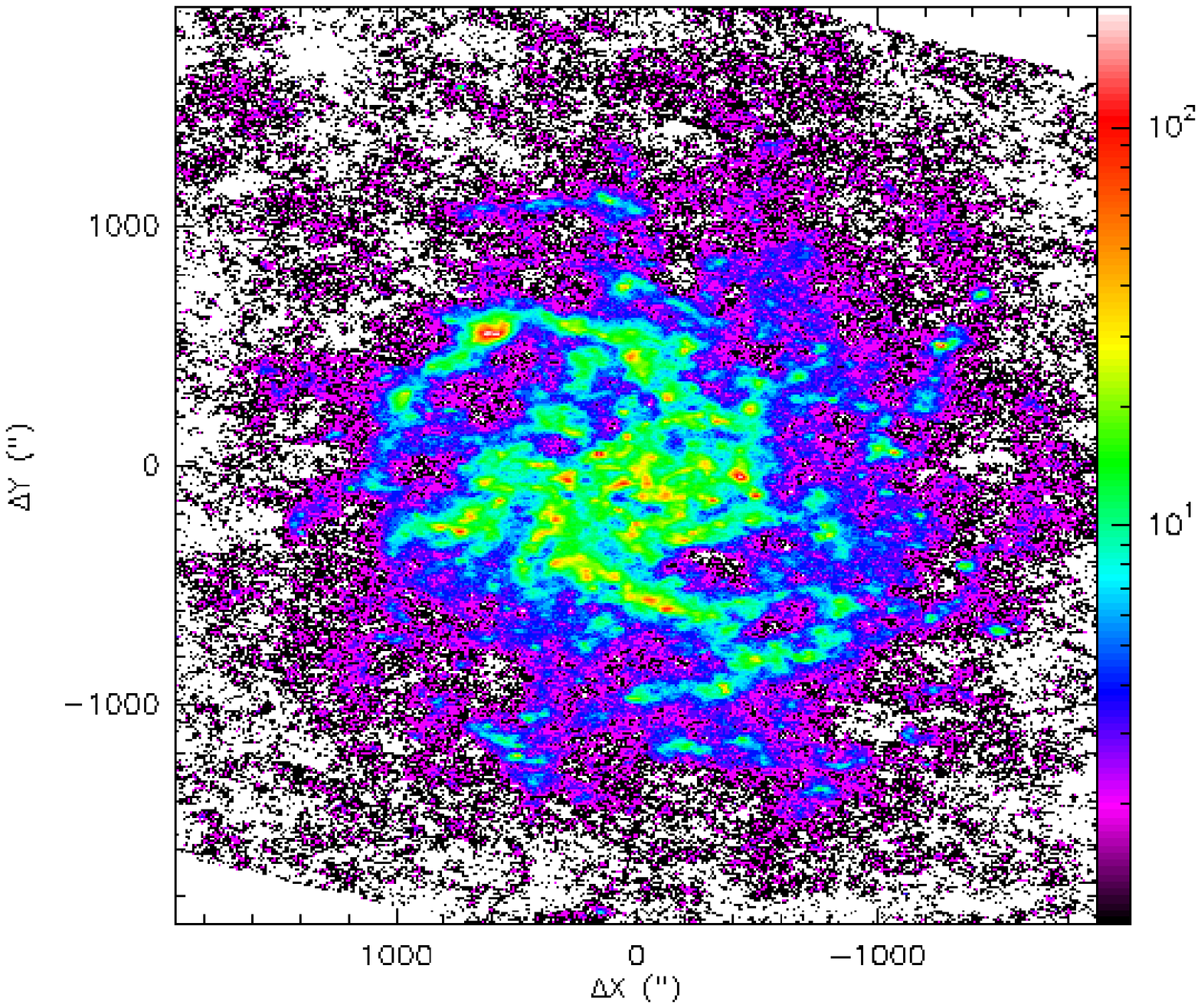}
\caption{Deprojected maps of the dust at 100 (left) and 160\mics (right).
The major axis has been rotated to be vertical. The color scale is in arbitrary units.}
\label{fig:dep1}
\end{figure*}

\begin{figure*}
\includegraphics[angle=-90,width=8.4cm,clip=true]{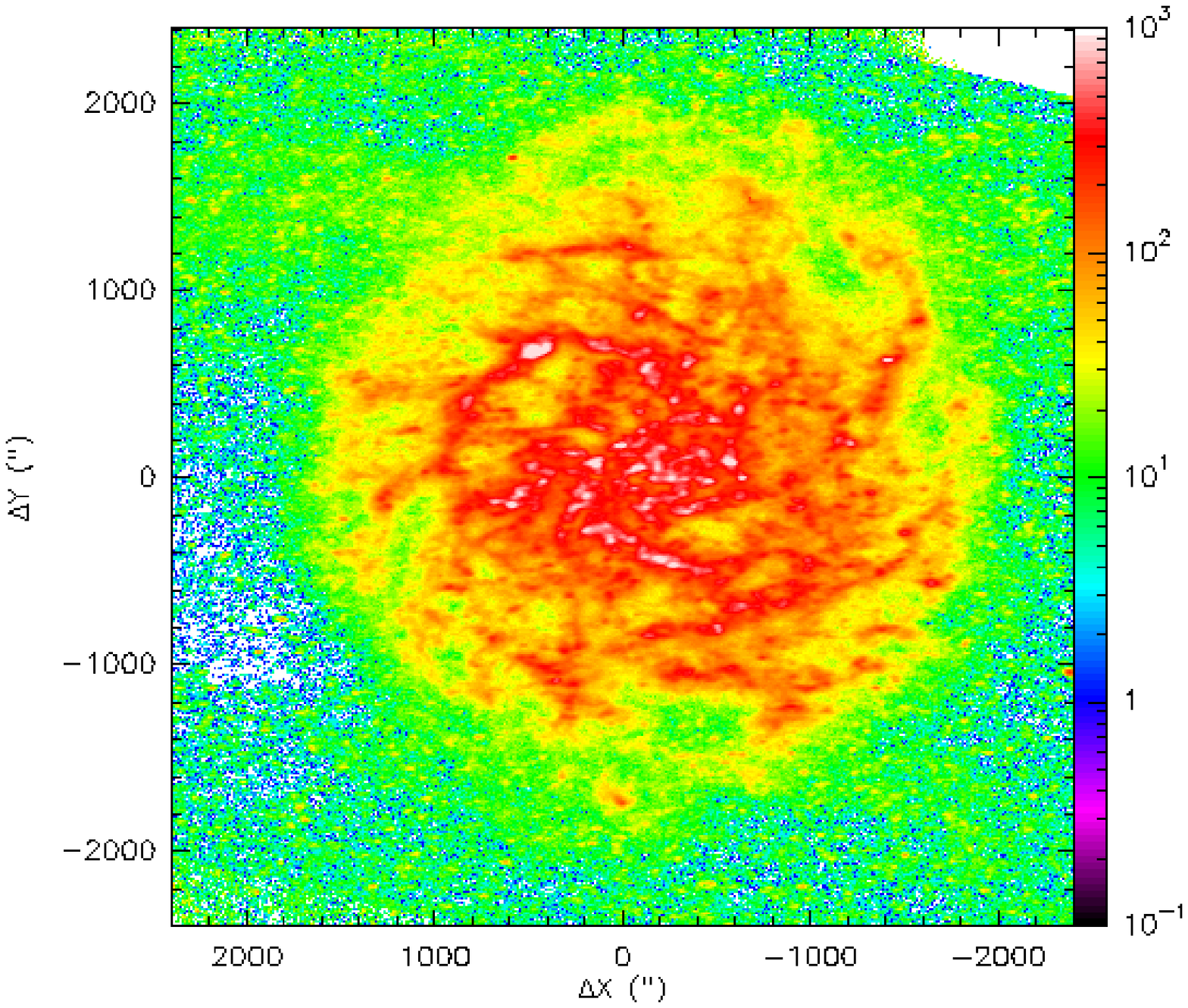}
\includegraphics[angle=-90,width=8.4cm,clip=true]{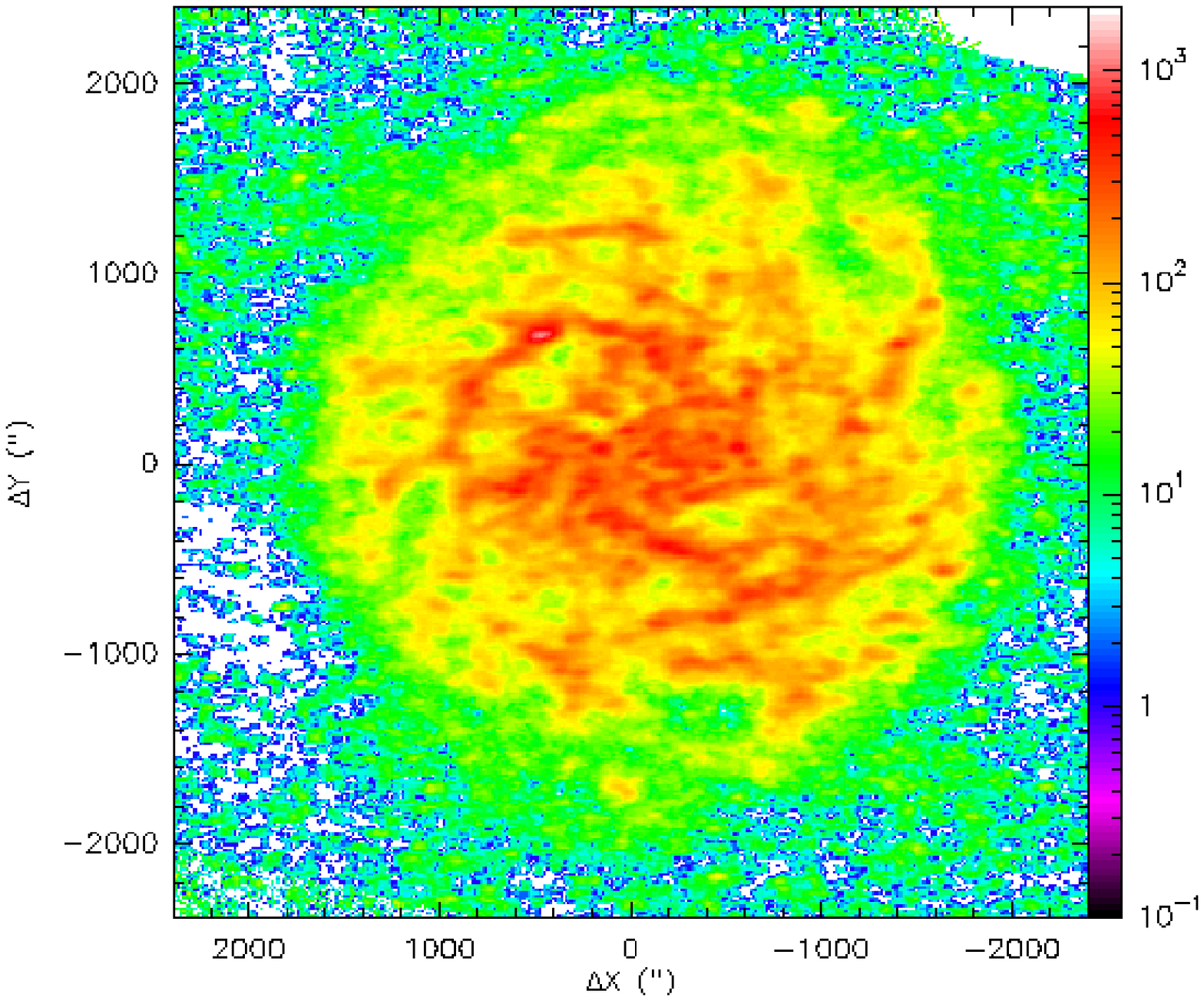}
\caption{Deprojected maps of the dust at 250 (left) and 500\mics (right). 
The major axis has been rotated to be vertical. The color scale is in arbitrary units.}
\label{fig:dep2}
\end{figure*}

An example of 2D-power spectrum is given for the PACS-100\mics image, in Figure \ref{fig:2D}.
The one-dimensional power-spectrum $P(k)$ can be calculated through azimuthal averaging,
in the ($k_x$,$k_y$) plane, using $k^2=k_x^2+k_y^2$. The power spectra
$P(k)$ are plotted for the various maps in Figures \ref{fig-pow1}, \ref{fig-pow2}, \ref{fig-pow3},
 \ref{fig-pow4}, \ref{fig-pow5} and \ref{fig-pow6}.
Two slopes have been fitted at  small (ss) and large scales (LS), and the different values
can be found in Table  \ref{tab:ps1}, together with the break scale separating
the two regimes.
In most of the cases, the power spectra show clearly two different 
regimes, with different slopes, and small and large-scales,
for scales larger than the spatial resolution. A knee is apparent between
the two regimes (ss and LS), and indicates the break scale. 
 We have used a least-square-fit program to determine the slopes
of the two power-laws, and left open the position of the separation
between the two straight lines. 
 Varying the value of the separation, we select for the break
the value which minimises the sum of the $\chi^2$ of the two line fits (ss and LS).
In general the minimum is quite clear.
There is one ambiguous case, however, which is the
HI map, in Figure \ref{fig-pow6}. The two slopes are close, and the determination
less trustworthy.
Examples of the deprojected images of the {\em Herschel} dust emission are
displayed in Figures \ref{fig:dep1} and \ref{fig:dep2}. In addition
to the power spectra, the azimuthal large-scale structure
develops spiral arms, that are quantified and studied in the Appendix  \ref{aplot}. 
 The determination of the break scale is obviously more secure when it is found 
significantly larger than the resolution scale of the observations. 
We have tested the effect of spatial resolution by smoothing the PACS-100\mics map
in Appendix \ref{smooth}. In the observed power spectra
(e.g. Figures \ref{fig-pow1}-\ref{fig-pow5}), the limit of the instrumental resolution is
clearly seen, through the sudden change from a straight line to a dropping curve, at large $k$.
We have indicated with two vertical dashed lines, the value of the resolution scale on the 
major axis, and also the geometric mean of the resolution on both minor and major axis, since 
the de-projection to a face-on galaxy implies a factor 1/cos(56$^\circ$) = 1.79 larger beam on
the minor axis.  We conclude from Appendix \ref{smooth} that break scales at least twice larger   
than the beam scale are robust with respect to resolution effect.

We have also tested the robustness of our determinations by computing 
the power spectra using only one side of the galaxy. The easiest way
to compute that, with the same algorithm, was to remove in the map one half of the galaxy, say
the southern part, and complete the map by extending the northern part by point
symmetry through the center. For all maps, the different slopes and break scales
for the two sides of the galaxy are also displayed in Table  \ref{tab:ps1}:
the last columns display the extreme values found for the slopes
and breaks. This procedure maximizes the error bars, especially
in the LS-slopes, since the Norhern part of M33 is quite different from the Southern part.
 The slopes are determined with an error bar of the order of $\pm 0.2$ in general,
and the break scales are even more robust, within 15\%.
 We have also checked the effect of smoothing on the maps. As expected,
the power-spectrum at large-scale is unchanged, and only the power at the
extreme of the small-scales of the spatial resolution drops.  The break scale
also is enlarged by the resolution, when it is too close, and we can see for instance that
the break at 70\mics is larger than at 24 and 100\mics (Table \ref{tab:ps1}) only because of the coarser
resolution. The fact that some
maps have a pixel size smaller than the Nyquist size (half of the nominal
resolution) does not influence the results. We have truncated the power spectra
at the Nyquist sampling, in the figures.

Looking at Tables \ref{tab:maps} and \ref{tab:ps1}, it is ovious that the
break scale is affected by the spatial resolution of the maps.
Five maps have a ratio between break and resolution scales lower or equal to 3
(NUV, the 3 SPIRE, and CO), for the others, the ratio is
3.8 (FUV), 12.5 (H$\alpha$), 4.2 (MIPS-24\mic), 3.9 (MIPS-70\mic),
3.9 (PACS-100\mic), 3.6 (PACS-160\mic) and 3.3 (HI). For those, the 
determination of the break appears significant. As for the slope determination,
the large-scale ones are not affected, and their variation is meaningful:
the LS-slopes are flatter for the star-formation tracers than for the dust and gas,
and they are growing steeper with wavelength.

It is interesting to note that the H$\alpha$ map, 
tracing gas ionized by the UV photons from massive stars
and, indirectly, current star formation,
has power spectrum slopes significantly
flatter than
 the other power spectra of the ISM components. This
might be expected, since star formation gives more power to the small scales.
Its break scale, related to the plane thickness is higher.
The GALEX FUV and NUV,
tracing the recent, but past, star formation, have slopes more comparable
to the other ISM slopes however, and have smaller break scales, 
implying thinner layers. The break scale in H$\alpha$ 
is larger than in the UV because  H$\alpha$ traces the ionised gas,
and only indirectly the young stars. The ionised layer may be 
thicker, affected by bubbles, filaments, fountain effects associated
to star formation.

%--------------------------------------Five column Table-----------------1-
\begin{table}[h]
      \caption[]{Power-law slopes and breaks for observations}
         \label{tab:ps1}
            \begin{tabular}{l c c c c c c }
            \hline
            \noalign{\smallskip}
  Map & slope  & slope & break &  \multicolumn{3}{c}{Extreme values} \\
       & LS  &   ss & (pc) & LS & ss & (pc) \\
            \noalign{\smallskip}
            \hline
            \noalign{\smallskip}
FUV & -1.2 & -3.4 & 44 &  -0.9/-1.2& -3.4/-3.9& 40-45\\
NUV & -1.2 &  -3.6 & 56 &  -1.0/-1.2& -3.0/-3.6& 50-56\\
H$\alpha$ & -.77 &  -2.4 & 300 &  -0.7/-1.0& -2.4/-2.7& 250-300\\
24\mic & -.9 &  -2.6 & 93 & -0.9/-1.0& -2.6/-2.7& 93-105\\
70\mic &  -1.5 &  -3.5 & 236 & -1.4/-1.5& -3.4/-3.8& 230-270 \\
100\mic & -1.5 &  -3.4 & 93 &   -1.4/-1.5& -3.2/-3.6& 93-105\\
160\mic & -1.5 &  -3.6 & 165 &  -1.5/-1.6& -3.6/-4.0& 140-165\\
250\mic & -1.6 & -4.7 & 208 & -1.4/-1.6& -4.3/-4.7& 200-210\\
350\mic & -1.3 &  -4.0 & 315 & -1.3/-1.5& -4.0/-4.4& 290-320\\
500\mic & -1.1 &  -4.4 & 390 & -1.1/-1.4& -4.2/-4.6& 320-390\\
CO(2-1) & -1.5 &  -3.7 & 103 &  -1.3/-1.5& -3.6/-3.8& 100-115\\
HI & -2.4 &  -3.2 & 110 & -2.0/-2.4& -3.2/-3.6& 100-120\\
%HI-5sec & -1.2 &  -- & -- &   No break\\
            \noalign{\smallskip}
            \hline
           \end{tabular}
\\LS= Large Scale ($>$ 500 pc), ss= small scale ($<$ 500 pc)
\end{table}

%--------------------------------------Five column Table-----------------2-
\begin{table}[h]
      \caption[]{Power-law slopes and breaks for simulations}
         \label{tab:ps2}
            \begin{tabular}{l c c c c c}
            \hline
            \noalign{\smallskip}
  Run & slope  & slope & break & Comments & $H_z$ \\
       & LS  &   ss & (pc) &     &  (pc)  \\
            \noalign{\smallskip}
            \hline
            \noalign{\smallskip}
run1 & -0.8 &  -3.6 & 75 & low feedback &80\\
run2 & -0.9 &  -3.4 & 75 &    $''$   &80\\
run3 & -2.3 &  -4.9 & 650 & extreme feedback& 217\\
run4 & -1.3 &  -3.8 & 63 & fiducial & 120\\
run5 & -2.1 &  -5.0 & 300   & more SF (C*) & 142\\
run6 & -1.1 &  -4.0 & 430 & higher threshold & 103\\
run7 & -0.8 &  -2.8 & 75 & no SF & 64\\
run8 & -0.9 &  -2.9 & 201 & no SF, Vs=12 \kms & 110\\
run9 & -1.1 &  -5.2 & 250 & SF, Vs=12 \kms & 108\\
            \noalign{\smallskip}
            \hline
           \end{tabular}
\\LS= Large Scale, ss= small scale
\\ The gas scale height $H_z$ = 2.35 $(<z^2>-<z>^2)^{1/2}$
\end{table}

The dust emissions at various wavelengths have quite compatible power spectra.
Their different break scales reflect the increasing beam size
with wavelength. As for the gas, the molecular medium appears thinner
than the atomic one, although spatial resolution might influence
its break scale.

Independent of the spatial resolutions, the slope of the power-spectra
at small scale are steeper at large wavelengths for the dust emission.
Between 250 and 500\mic, the SPIRE maps have slopes between -4.3 and -4.7,
while the MIPS and PACS maps between 24 and 100\mics have slopes
comprised within -2.7 and -3.8.  This is significant, and tends to show
that the cool dust has less small-scale structure, since individual cloud
cores are invisible.

\subsection{Power-spectrum of the model}
\label{pow-mod}

\begin{figure}[htp]
\includegraphics[angle=-0,width=8.5cm,clip=true]{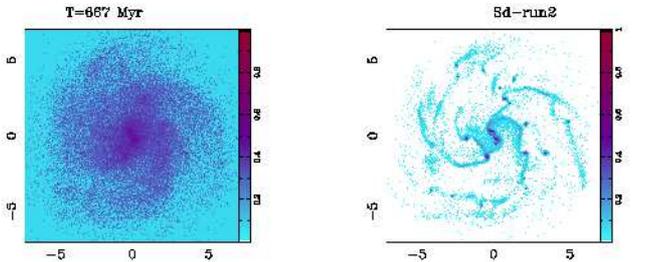}
\caption{Face-on projection of the run2 model, at T=667 Myr. {\bf Left} are the stars,
and {\bf Right}, the gas. The linear scales on both axes are in kpc,
and the color scale is in arbitrary units.}
\label{fig:run2}
\end{figure}

\begin{figure}[htp]
\includegraphics[angle=-0,width=8.5cm,clip=true]{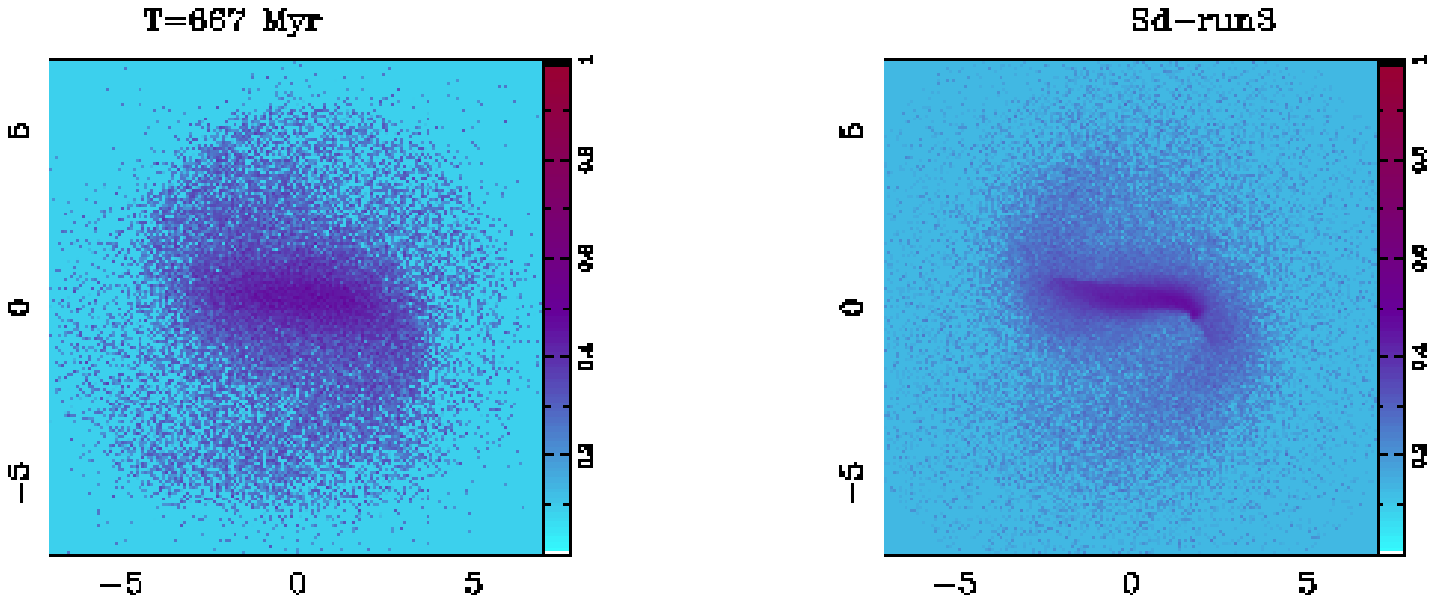}
\caption{Face-on projection of the run3 model, at T=667 Myr. {\bf Left} are the stars,
and {\bf Right}, the gas. The linear scales on both axes are in kpc,
and the color scale is in arbitrary units.}
\label{fig:run3}
\end{figure}

\begin{figure}[htp]
\includegraphics[angle=-0,width=8.5cm,clip=true]{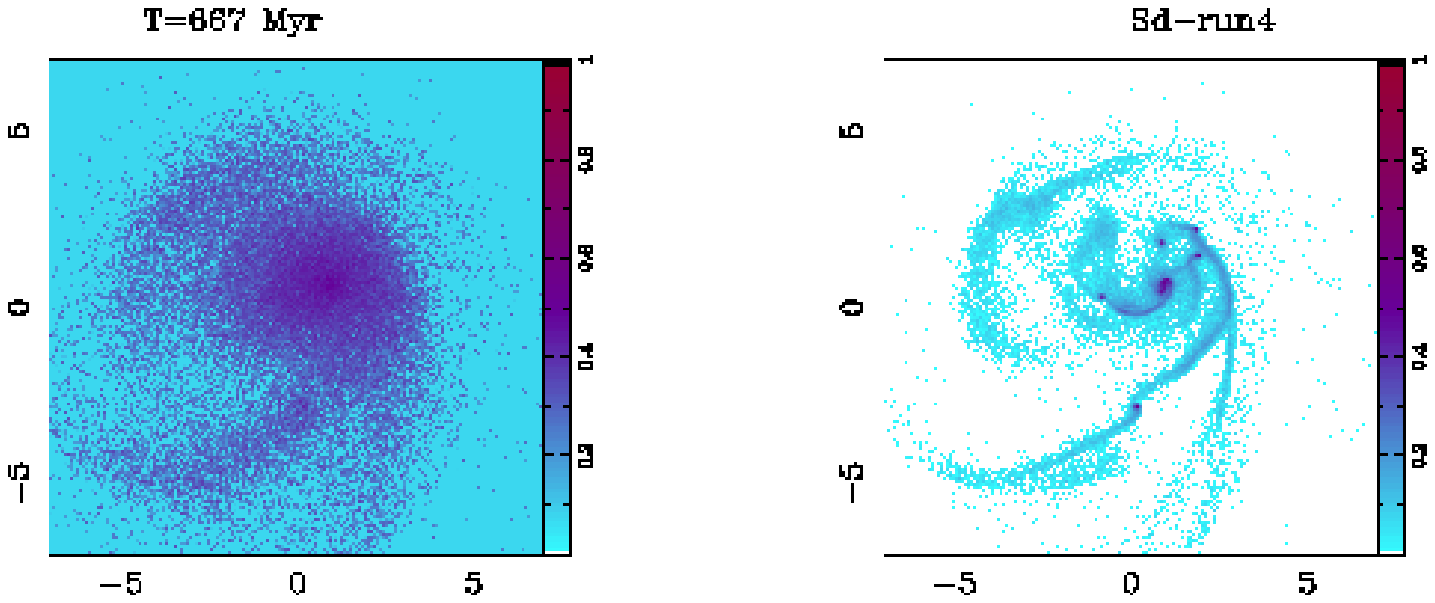}
\caption{Face-on projection of the run4 model, at T=667 Myr. {\bf Left} are the stars,
and {\bf Right}, the gas. }
\label{fig:run4}
\end{figure}

\begin{figure}[htp]
\includegraphics[angle=-0,width=8.5cm,clip=true]{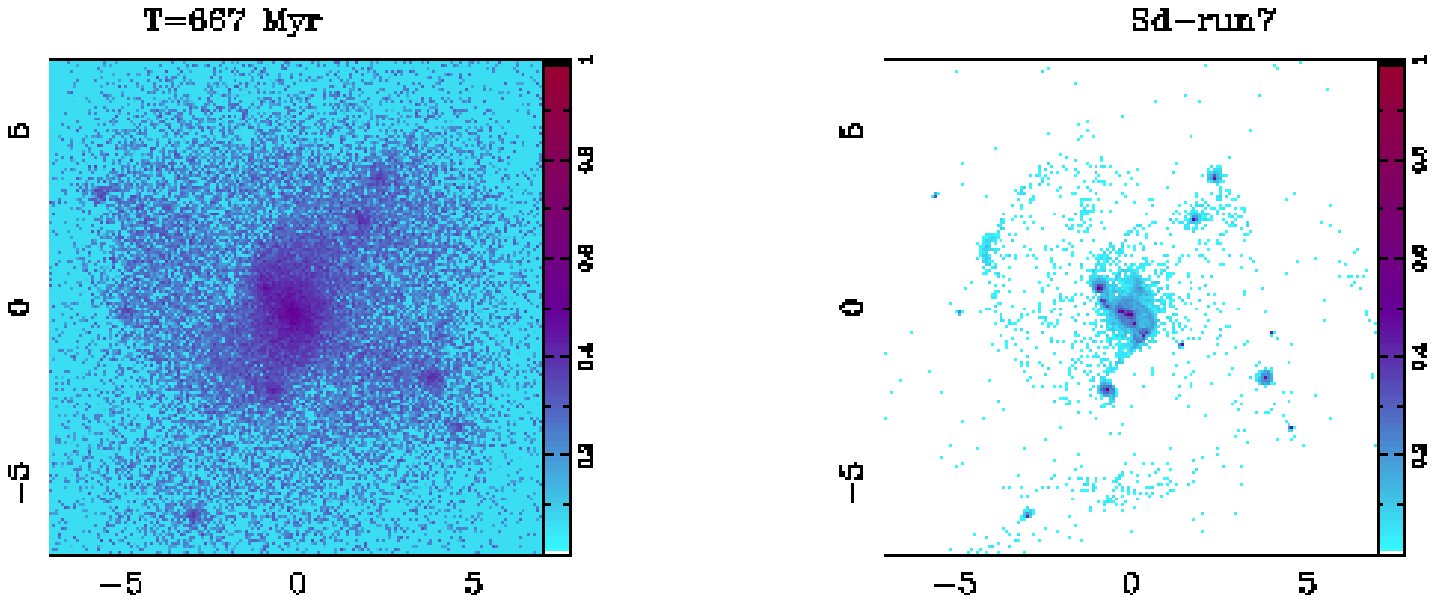}
\caption{Face-on projection of the run7 model, at T=667 Myr. {\bf Left} are the stars,
and {\bf Right}, the gas. }
\label{fig:run7}
\end{figure}

\begin{figure}[htp]
\includegraphics[angle=-0,width=8.5cm,clip=true]{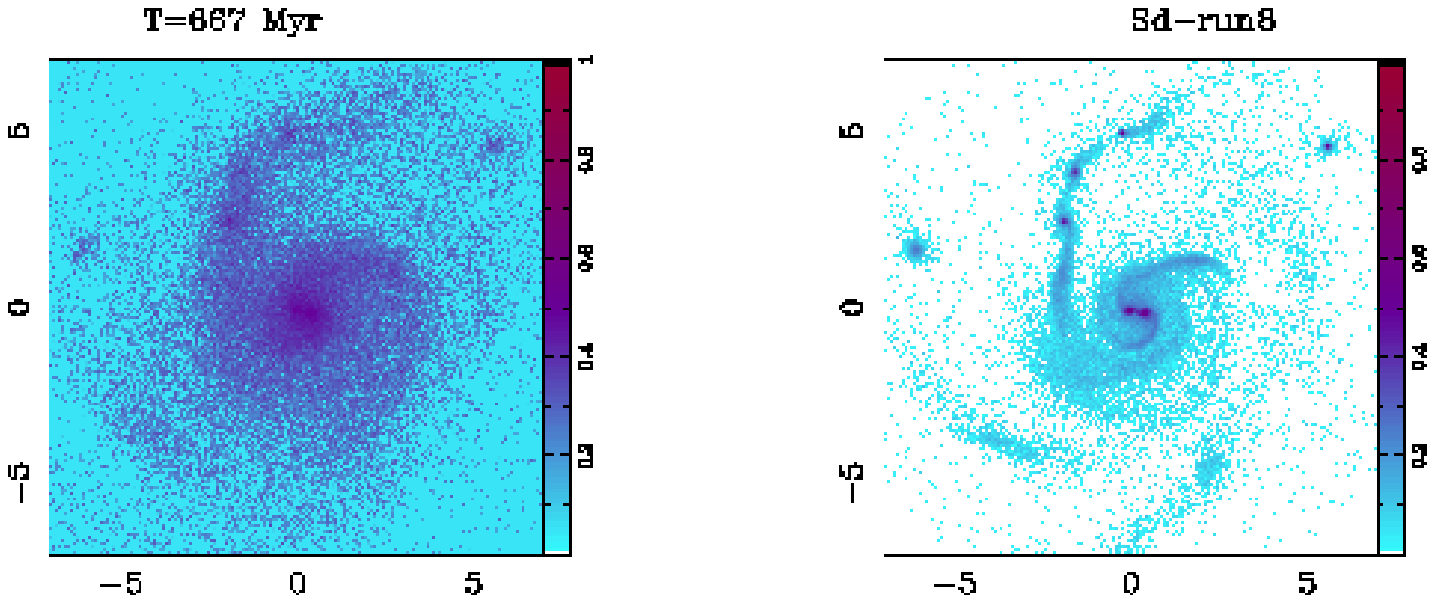}
\caption{Face-on projection of the run8 model, at T=667 Myr. {\bf Left} are the stars,
and {\bf Right}, the gas. }
\label{fig:run8}
\end{figure}

\begin{figure}[htp]
\includegraphics[angle=-0,width=8.5cm,clip=true]{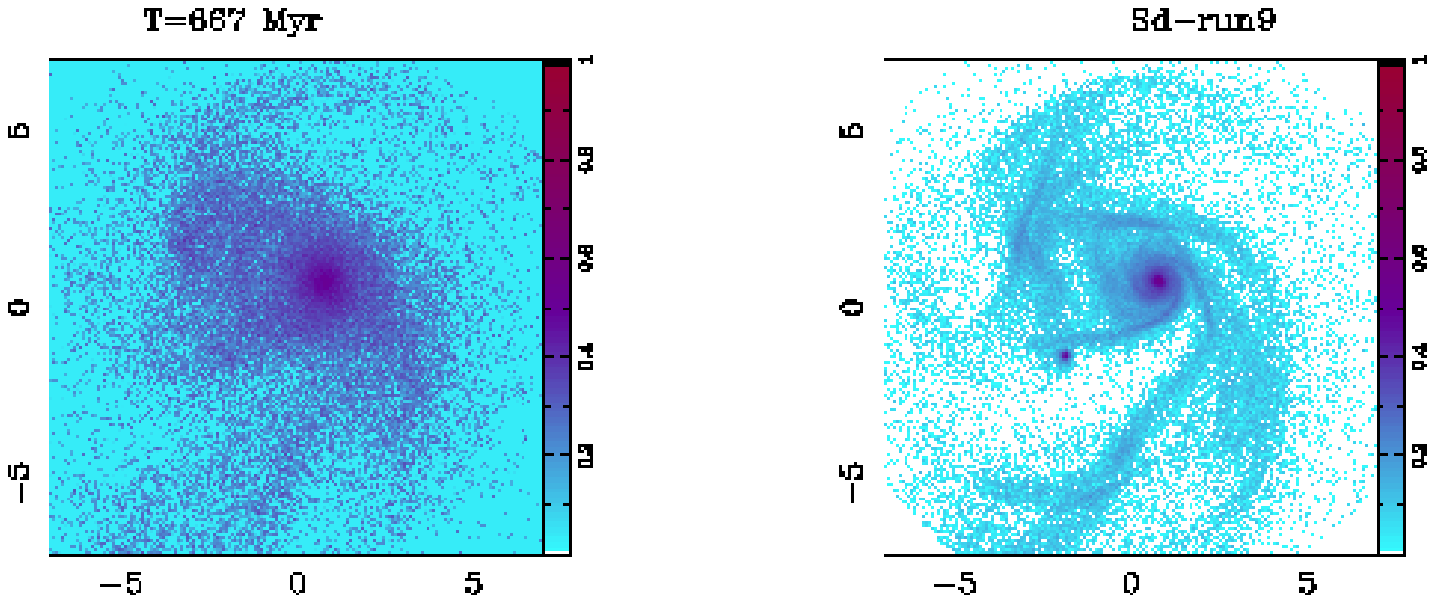}
\caption{Face-on projection of the run9 model, at T=667 Myr. {\bf Left} are the stars,
and {\bf Right}, the gas. }
\label{fig:run9}
\end{figure}

 The various runs develop different gas structure and clumpiness,
as can be seen in Figures \ref{fig:run2} to  \ref{fig:run9}, due to both initial conditions
that are selected more or less stable, or with different sub-grid ``temperature'' of
the gas, and to the star formation rate and feedback. Some galaxies, which are
launched with initial conditions among the most stable, 
can end up with a more perturbed morphology and larger turbulence
than others started more unstable, due to a larger supernovae feedback.

\begin{figure}[htp]
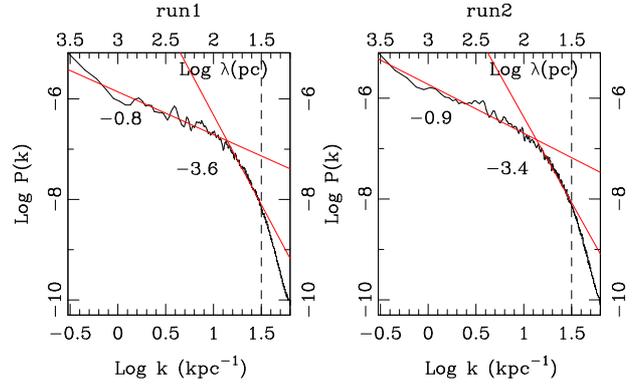

\includegraphics[angle=-90,width=4cm]{18282f17a.ps}
\includegraphics[angle=-90,width=4cm]{18282f17b.ps}
\caption{Power spectrum of the gas in models run1 and run2. The vertical dash line represents 
the softening length of the gravity. }
\label{fig-powm1}
\end{figure}

\begin{figure}[htp]
\includegraphics[angle=-90,width=4cm]{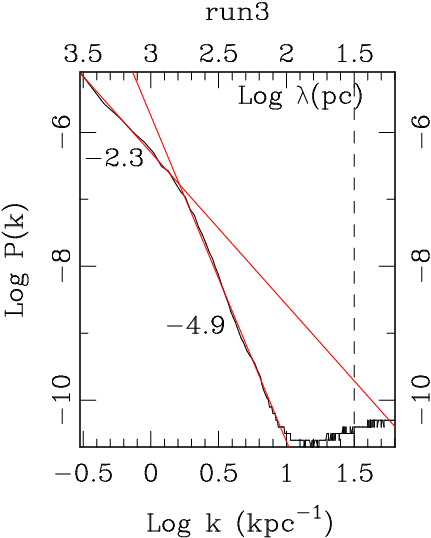}
\includegraphics[angle=-90,width=4cm]{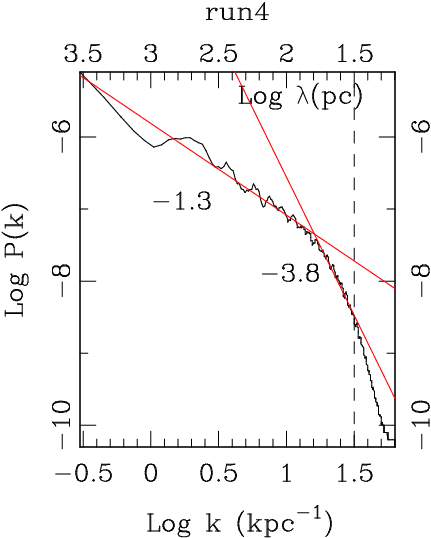}
\caption{Power spectrum of the gas in models run3 and run4.  }
\label{fig-powm2}
\end{figure}

We computed the Fourier analysis in the same way on the gas maps
obtained in the numerical simulations.  The derived results are converging 
towards the same values at the middle of the runs, and we plot all of them
for the final snapshots at T= 667 Myr for the sake of simplicity
(cf Figures \ref{fig-powm1}, \ref{fig-powm2} and \ref{fig-powm3}).
 For comparison, we also plot the power-spectrum of some old stellar components,
which have almost only one power-law (Figure \ref{fig:runst}). These correspond to run 2 and run 7 at T= 667 Myr.
There is structure until about 300 pc, and no small-scale component, as expected for a
collisionless component.

The first result which is obvious in Table \ref{tab:ps2} is that the
predictions of the models bracket the values obtained in the observations,
both in the slopes at large and small scales, and in the break scales.
Those can be compared to the actual thickness of the gas layer, in the simulations.
This is done in Table \ref{tab:ps2}, showing that the break scale
does not follow closely the thickness. 
Although both should be of the same order when only the self-gravity of the gas
is taken into account, their close relation is blurred by other factors,
and in particular the feedback efficiency.
When the feedback is too extreme,
the small scale structure of the gas is smoothed out, even towards larger scale
than the vertical thickness. So the range of break scales is wider
than that of possible plane heights.

 The main  parameters that control the turbulence and
therefore the power-spectrum of the gas structure are the star formation rate, 
the amount of feedback, but also the assumed equivalent temperature of the gas,
or sound speed, directly involved in the pressure forces. Increasing the turbulence
by supernovae feedback suppresses efficiently the small-scale structure
(power-law slopes of $\sim$ -5.), and thickens substantially the gas layer,
and the corresponding break scale. 
Without star formation and feedback, it is only possible to have gas
morphologies comparable to the observed ones if
a minimum sound speed of 12 \kms is injected.
 Figures \ref{fig:run2} to \ref{fig:run9} show the large
variety of gas structures, according to the variations of those
parameters.
 It appears that our fiducial model (run4) is the one which corresponds
the best to the M33 galaxy, both for its morphology, and for the typical
power-law slopes and break scale. It has a star formation rate of
$\sim$ 0.7 M$_\odot$/yr, and a feedback such that 10\% of supernovae energy
is coupled to the kinematics of the gas. Verley \etal (2009) found for M33
a star formation rate of 0.5 M$_\odot$/yr.

\begin{figure}[htp]
\includegraphics[angle=-90,width=4cm]{18282f19a.ps}
\includegraphics[angle=-90,width=4cm]{18282f19b.ps}
\caption{Power spectrum of the gas in models run7 and run8.  }
\label{fig-powm3}
\end{figure}

\begin{figure}[htp]
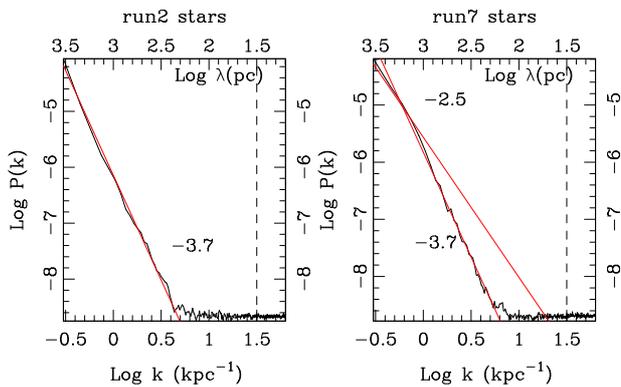

\includegraphics[angle=-90,width=4cm]{18282f20a.ps}
\includegraphics[angle=-90,width=4cm]{18282f20b.ps}
\caption{Power spectrum of the stellar component in run2 and run7. The vertical dash line represents 
the softening length of the gravity. }
\label{fig:runst}
\end{figure}

\section{Discussion and conclusions}
\label{disc}
We have presented the Fourier analysis of dust and gas emission maps of M33 at several wavelengths,
and compared them with the H$\alpha$  and UV maps, tracers of recent star formation.
  We find that all the power spectra can be decomposed into two regimes,
a large-scale ($>$ 500 pc) power-law, of $-1.5<$ slope$_{LS} <-1.0$, and a steeper small-scale one, 
with $-4.0<$ slope$_{ss} <-3.0$. The latter slopes correspond quite well to what is
found for the Milky Way (Dickey \etal 2001) or the Small Magellanic Cloud (Stanimirovic \etal 2000).
The H$\alpha$ star formation tracer has shallower power-spectra, indicating
more power at small scales. The break scale is of the order of 100-150 pc for the dust and the gas,
increasing at large wavelength, probably influenced by the progressive lack of spatial resolution.
  The H$\alpha$ break scale is significantly higher (300 pc), implying a thicker layer for ionized gas,
while the UV layers appears thinner (65 pc).
This might be understood considering that
UV originates from stars that are concentrated in the disk whereas 
the ionised gas could be present quite far from the plane of the disk:
giant HII regions have typically round shapes, and are not as 
confined to the disk.
Tabatabaei \etal (2007) have already shown with wavelets
that the H$\alpha$ map is dominated by giant HII regions, which are bubble-like
with sizes from 100 to 500 pc.   By comparison the diffuse component
is weaker, which flattens the Fourier ss-slope, and enlarges the break scale.
Although the warm dust in the MIPS-24\mics map is also related to the young star formation,
there are no such large bubbles: only the point sources are prominent inside
the bubble, as can be seen in Figure 3 of Verley \etal (2010). This explains the different break scales between 
H$\alpha$ and MIPS-24\mics maps.

It is also possible that the large thickness suggested by the H$\alpha$ analysis is related to the
diffuse ionized gas (DIG) observed in several edge-on galaxies to follow a rather thick layer
of about 1 kpc height (see a review by Dettmar 1992). This extra-planar ionized gas is thought
to be due to star formation, it is present in about 40\% of 74 edge-on galaxies (Rossa \& Dettmar 2003).
It can be widely diffuse, or take the shape of filaments, plumes and bubbles (Rand, 1997).
These extra-planar extensions could be due to UV ionizing photons escaping the star forming regions,
supernovae feedback, fountain effects, chimneys,
and be part of an active disk-halo interaction, with exchange of matter (Howk \& Savage 2000).

  The comparison with simulated models tells us that the observation of a break in the power-spectrum
of observed gas distribution is not always related to the 2D/3D transition, and indicates the
thickness of the gaseous disk only when the latter is not heated by a strong star formation
rate and a strong supernovae feedback. For a quiescent galaxy, the break should still
be an indication of the gas layer thickness.

Block \etal (2010) found in the LMC that the small scale slopes of the power spectra were steeper at longer
wavelengths, i.e. the cool dust was smoother than the hot emission. We find a similar result.
The slope of the SPIRE maps are around -4.5, while the warmer dust at 24-100\mics has 
an average slope of -3.2.  
 This observation can be interpreted in terms of star formation occuring in the clumpy medium.
 Once the stars are formed, they heat the dust around, and this explains why the emission 
at shortest wavelength (warmer dust) is more clumpy.  Also the break scale, and thus
the thickness of the plane, is larger for the cold gas. This could be due in addition to the fact that 
the cold component is much more extended in radius (Kramer \etal 2010), and it is well known that the gas layer
is flaring in the outer parts of galaxies. 
As for the gas emission, the molecular gas appears more clumpy,
but only slightly more so than the atomic gas.  And the break occurs at about the same scale.
Since it is likely that in the relatively quiescent M33 galaxy, the break indicates the thickness of the gas
layer, this situation is not alike the Milky Way, where the CO component is significantly thinner
than the HI (Bronfman \etal 1988). A thick CO layer, 3 times as wide as the dense molecular gas layer, and
comparable to the HI layer, has been observed (Dame \& Thaddeus 1994), but it involves a small mass.
M33 being a galaxy of intermediate mass, could have indeed a thicker gas layer
than the Milky Way (Dalcanton \etal 2004).

We also showed that numerical models predict gas morphologies and small scale
power-laws which bracket the observational data. It is possible to obtain a 
colder or a more turbulent interstellar medium, by varying the star formation
feedback and the sound speed of the gas.
 The Fourier analysis of the models reveals that the break scale is a good
indicator of the actual thickness of the gas layer, although
it varies within a wider range when extreme feedback is used, and smoothes
out the small scale gas structures. The power-law slopes obtained in the models
at large and small scale reflect quite accurately the gas morphology and turbulence.
 Fourier analyses are therefore a good tool to 
determine the thickness of the gas plane, at any orientation on the plane of the sky,
and the actual physical state of the gas, in its star formation phase.
Note however that the current models are not devoid of shortcomings: they only partly
allow to distinguish between the 12 different tracers of the gas phase
observed here, as radiative transport, dust properties, etc. are
not taken into account.

 Comparison of power spectrum analysis between galaxies should allow a 
determination of their relative star formation rate and feedback. This will need however
high spatial resolution maps, such as the ones that will be provided with ALMA.

 The gas plane in M33 appears relatively thick and turbulent, with respect
to our closest fiducial model, which could be
a consequence of its recent heating in particular through the M31 interaction
(McConnachie \etal 2009).
% Warp, flare? 

%%%%%%%%%%%%%%%%%%%%%%%% acknowledgments
\begin{acknowledgements}
  We warmly thank the anonymous referee for his/her constructive comments
and suggestions. 
We have made use of the NASA/IPAC Extragalactic Database (NED).
\end{acknowledgements}
%%%%%%%%%%%%%%%%%%%%%%%%%%%%%%%%%%%%%

\appendix
\section{Fourier analysis of the density}
\label{aplot}
 Until now, we have essentially analysed the small scale structure of the interstellar medium
in various tracers, but ignored the azimuthal dependence.
 The different tracers also reveal a contrasted spiral structure, that is
interesting to present, at least in two representative deprojected maps,
the {\em Herschel} 100\mics image (dust and gas images show similar features), 
and the H$\alpha$ image (star formation, similar to UV images).

We have computed the Fourier transform of the corresponding
surface densities, decomposed as:

$$
\mu(r,\phi) = \mu_0(r) + \sum_m a_m(r) \cos (m \phi - \phi_m (r))
$$

where  the normalised strength of the Fourier component $m$ is
$A_m(r) =  a_m(r) / \mu_0 (r)$. Thus, $A_2$ represents the normalized amplitude 
for a $m=2$ distortion like a bar or spiral arm, at a given radius, and 
$A_1$ represents the same for the disk lopsidedness.
The quantity $\phi_m$ denotes the position angle or
the phase of the Fourier component $m$. Such a density analysis
has been made for example by Bournaud \etal (2005), to study asymmetries
and lopsidedness in galaxies, and correlate it with $m=2$ features.

Figure \ref{fig:aplot-100} shows a plot of the amplitude $A_m (r) $ and the phase
$\phi_m (r)$ versus radius $r$ for M33, at 100\mics and 
Figure \ref{fig:aplot-Ha}, the same for H$\alpha$.
Both show a prominent spiral feature at radii $\sim$ 3-4 kpc,
with a phase decreasing slightly with radius, in the direct sense.
 The $m=2$ distortions show accompanying $m=4$ harmonics.
The lopsidedness also increases with radius, while the 
the corresponding phase is fairly constant.

These plots show that the spiral structure in M33 is
mildly contrasted, in both gas and SFR tracers. Indeed, the normalised
$m=2$ amplitude is between 0.4 and 0.8. 
There is also a significant asymmetry and lopsidedness, as testified by the strong
$m=1$ amplitude comparable with the $m=2$. 
This explains why the slopes at large-scale were more different between
the two halves of the galaxy than the small-scale ones.
This large asymmetry could be due to the current interaction between
M33 and M31 (McConnachie \etal 2009).

\begin{figure*}
\centering
   \includegraphics[height=17cm,angle=-90]{18282fA1.ps} \caption{
 {\bf Left}: Radial variations of A$_m$, for $m=1-4$, the normalised Fourier
amplitudes of the deprojected 100\mics image of M33.
 {\bf Right}: Corresponding radial variations of the phases $\Phi_m$ in radians.}
\label{fig:aplot-100}
\end{figure*}

\begin{figure*}
\centering
   \includegraphics[height=17cm,angle=-90]{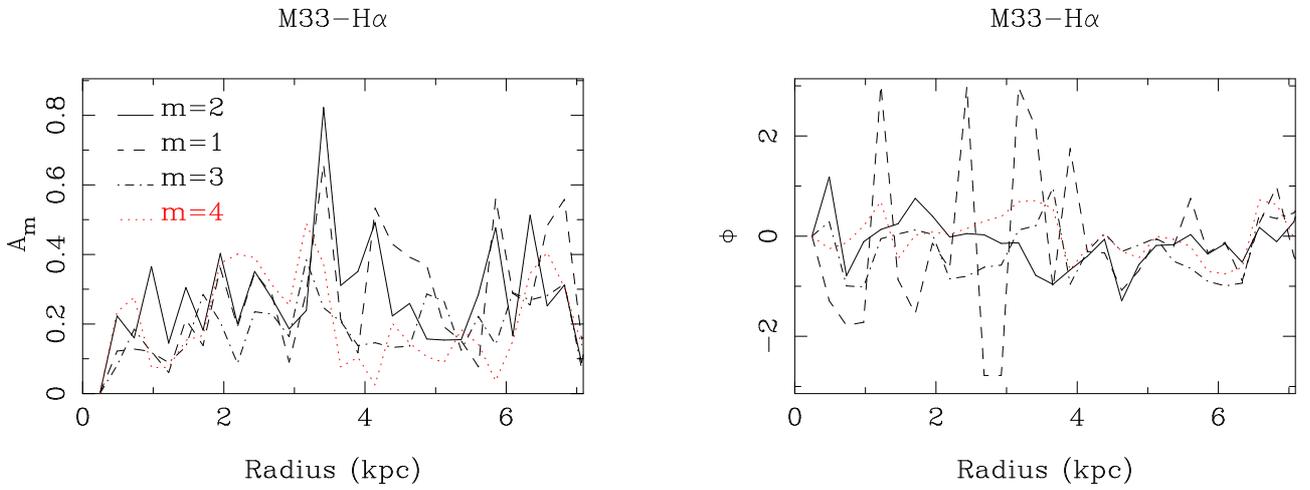} \caption{
 {\bf Left}: Radial variations of A$_m$, for $m=1-4$, the normalised Fourier
amplitudes of the deprojected H$\alpha$  image of M33.
 {\bf Right}: Corresponding radial variations of the phases $\Phi_m$ in radians.}
\label{fig:aplot-Ha}
\end{figure*}

\section{The effect of resolution}
\label{smooth}
 
\begin{figure}[htp]
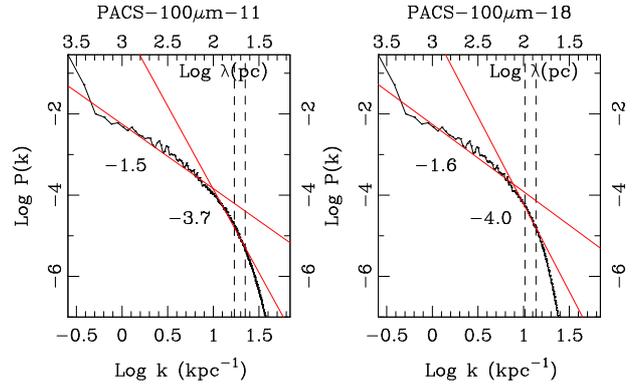

\includegraphics[angle=-90,width=4cm]{18282fB1a.ps}
\includegraphics[angle=-90,width=4cm]{18282fB1b.ps}
\caption{Power spectrum of the PACS 100\mics map,
at degraded resolutions of 11 and 18\arcsec.  The vertical dash line at right represents
the smoothed spatial resolution, on the major axis.
Since the de-projection to a face-on galaxy implies a factor 1/cos(56$^\circ$) = 1.79 larger beam on
the minor axis, we have also plotted a second dashed vertical line at left,
indicating the geometric mean of the resolution on both minor and major axis.
}
\label{fig:100-23}
\end{figure}

\begin{figure}[htp]
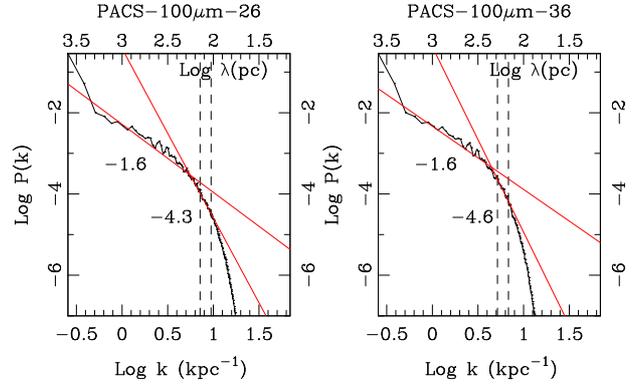

\includegraphics[angle=-90,width=4cm]{18282fB2a.ps}
\includegraphics[angle=-90,width=4cm]{18282fB2b.ps}
\caption{Same as Figure \ref{fig:100-23} for the
degraded resolutions of 26 and 36\arcsec.  
}
\label{fig:100-45}
\end{figure}

We have explored the effect of spatial resolution in the determination 
of the various slopes and break scales, in smoothing the PACS-100\mics map
at various resolutions, from the original 7\arcsec\ts to 11, 18, 26 and 36\arcsec,
corresponding to 44, 72, 104 and 144 pc on the major axis. 
 The resulting fits are shown in Figure \ref{fig:100-23} and  \ref{fig:100-45}, to be compared
with Figure \ref{fig-pow1}.
The break scales are respectively 97, 111, 135, 195 pc, instead of 93 pc for the 
original  7\arcsec\ts resolution.  The determination is possible only for the two
first cases.

\end{document}